\pdfoutput=1
\documentclass[11pt,twoside,a4paper,cmspaper,final,collab]{cms-tdr}
\def\svnVersion{4d6917e}\def\svnDate{2024/07/04}\def\cmsCernNoTag{CERN-EP-2024-020}\def\cmsCernDate{\today}\def\cmsMessage{Published in Physical Review Letters as \href{http://dx.doi.org/10.1103/PhysRevLett.132.261902}{\doi{10.1103/PhysRevLett.132.261902}.}}
\begin{document}\cmsNoteHeader{TOP-22-001}

\renewcommand{\cmsCollabName}{The ATLAS and CMS Collaborations}
\renewcommand{\cmsTag}{CMS-\cmsNUMBER\\&ATLAS TOPQ-2019-13\\}
\renewcommand{\cmsPubBlock}{\begin{tabular}[t]{@{}r@{}l}&CMS \cmsSTYLE~\cmsNUMBER\\&ATLAS TOPQ-2019-13\\\end{tabular}}
\renewcommand{\cmsCopyright}{\copyright\,\the\year\ CERN for the benefit of the CMS and ATLAS Collaborations.}
\renewcommand{\appMsg}{See Appendices~\ref{app:collab} and~\ref{app:atlas} for the lists of collaboration members}
\renewcommand{\cmslogo}{\includegraphics[height=2cm]{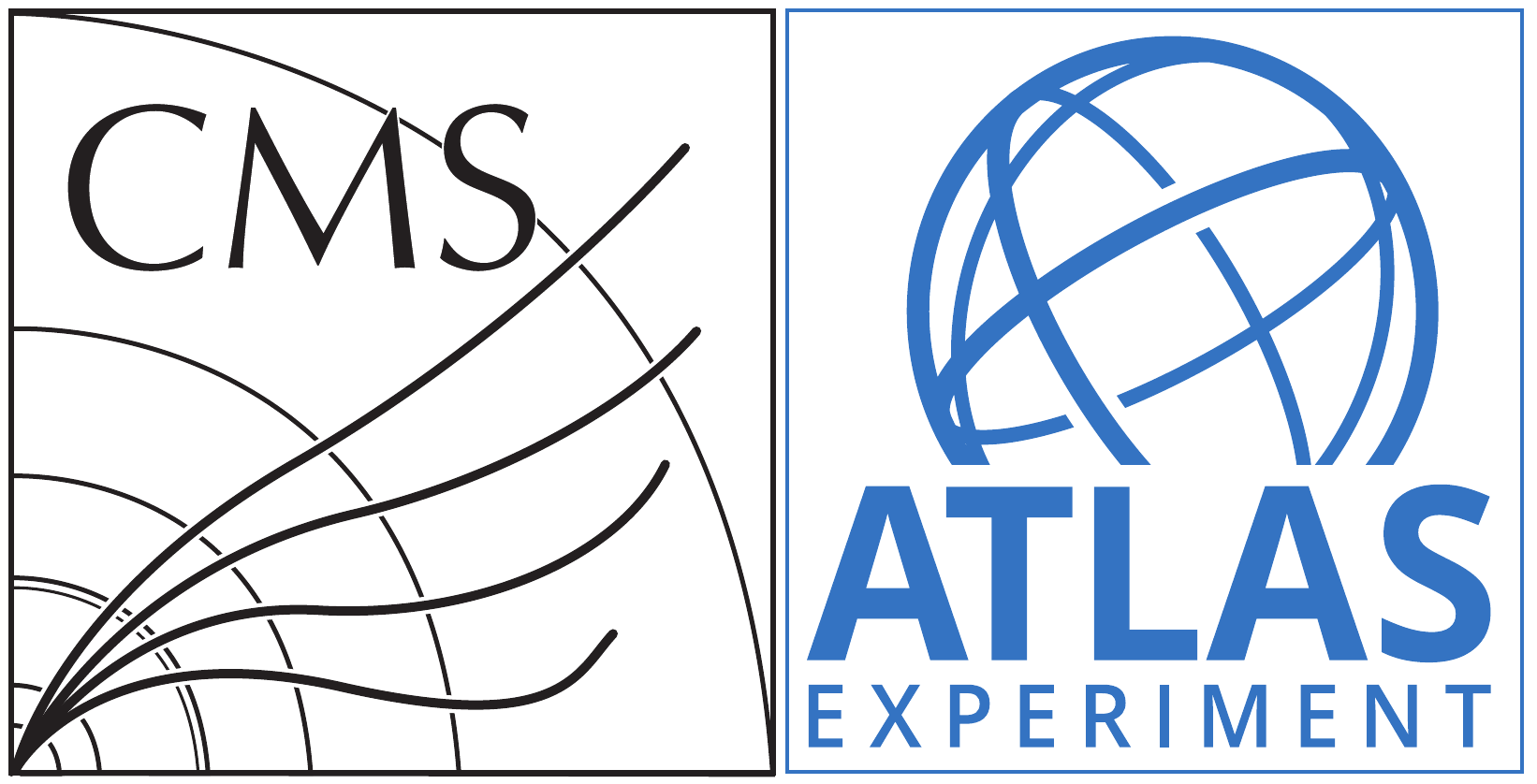}}
\ifthenelse{\boolean{cms@external}}{%
    \newcommand{\refsupplemental}{the Supplemental Material~\cite{this:Supplemental}\xspace}
}{%
    \newcommand{\refsupplemental}{Appendix~\ref{this:Supplemental}\nocite{placeholder}\xspace}
}

\providecommand{\cmsTable}[1]{\resizebox{\textwidth}{!}{#1}}
\newlength\cmsTabSkip\setlength{\cmsTabSkip}{1ex}

\newcommand{\mt}{\ensuremath{m_{\PQt}}\xspace}
\newcommand{\mtATLAS}{\ensuremath{\mt^\text{ATLAS}}\xspace}
\newcommand{\mtCMS}{\ensuremath{\mt^\text{CMS}}\xspace}
\newcommand{\mtLHC}{\ensuremath{\mt^\text{LHC}}\xspace}
\newcommand{\sigmt}{\ensuremath{\sigma_{\mt}}\xspace}

\newcommand{\LHCJES}[1]{\ensuremath{\text{\mbox{JES #1}}}\xspace}
\newcommand{\LHCfJES}[1]{\ensuremath{\text{\mbox{#1-JES}}}\xspace}
\newcommand{\LHCbJES}{\LHCfJES{\PQb}}
\newcommand{\LHCgJES}{\LHCfJES{\Pg}}
\newcommand{\LHClJES}{\LHCfJES{\Pl}}
\newcommand{\CMSJES}[1]{\ensuremath{\text{\mbox{CMS JES #1}}}\xspace}

\newcommand{\BR}{\ensuremath{\mathcal{B}}\xspace}
\newcommand{\hdamp}{\ensuremath{h_{\text{damp}}}\xspace}
\newcommand{\chisq}{\ensuremath{\chi^2}\xspace}

\cmsNoteHeader{TOP-22-001}
\title{Combination of measurements of the top quark mass from data collected by the ATLAS and CMS experiments at \texorpdfstring{$\sqrt{s}=7$ and 8\TeV}{sqrt(s)=7 and 8 TeV}}

\date{\today}
\abstract{A combination of fifteen top quark mass measurements performed by the ATLAS and CMS experiments at the LHC is presented. The data sets used correspond to an integrated luminosity of up to 5 and 20\fbinv of proton-proton collisions at center-of-mass energies of 7 and 8\TeV, respectively. The combination includes measurements in top quark pair events that exploit both the semileptonic and hadronic decays of the top quark, and a measurement using events enriched in single top quark production via the electroweak $t$-channel. The combination accounts for the correlations between measurements and achieves an improvement in the total uncertainty of 31\% relative to the most precise input measurement. The result is $\mt = 172.52 \pm 0.14 \stat \pm 0.30 \syst \GeV$, with a total uncertainty of 0.33\GeV.}

\hypersetup{%
pdfauthor={CMS Collaboration},%
pdftitle={Combination of measurements of the top quark mass from data collected by the ATLAS and CMS experiments at sqrt(s)=7 and 8 TeV},%
pdfsubject={ATLAS, CMS},%
pdfkeywords={ATLAS, CMS, top quark, top mass, LHC combination}}
\maketitle

The mass of the top quark (\mt) is a fundamental parameter of the standard model (SM).
Its precise measurement provides a crucial input to fits that probe the consistency of the SM~\cite{ALEPH:2010aa, Degrassi:2012ry, Bezrukov:2012sa, Bezrukov:2007ep, DeSimone:2008ei}.
The Tevatron experiments CDF and \DZERO were the first to measure \mt~\cite{CDF:1995wbb, D0:1995jca}, and produced a combined result in 2016~\cite{TeV:2016combination}.
During the 2011--2012 data-taking period of the CERN LHC, proton-proton collisions at $\sqrt{s}=7$ and 8\TeV produced large numbers of top quarks in pairs via strong interactions or singly via electroweak processes.
The two general-purpose experiments at the LHC, ATLAS~\cite{ATLAS:Detector-2008} and CMS~\cite{CMS:Detector-2008}, performed multiple measurements of \mt using these data~\cite{CMS:TOP-11-015, CMS:TOP-11-016, CMS:TOP-11-017, ATLAS:TOPQ-2013-03, ATLAS:TOPQ-2013-02, CMS:TOP-14-022, CMS:TOP-12-030, ATLAS:TOPQ-2016-03, CMS:TOP-15-014, ATLAS:TOPQ-2015-03, CMS:TOP-15-001, CMS:TOP-15-008, ATLAS:TOPQ-2017-03, CMS:TOP-23-003}.
In this Letter, a combined \mt measurement from the ATLAS and CMS experiments is published for the first time.
The 15 input measurements utilize up to 5 (20)\fbinv of integrated luminosity per experiment at 7 (8)\TeV.
A detailed estimate of the correlations between the ATLAS and CMS measurements is performed and the measurements are combined using the best linear unbiased estimate (BLUE) method~\cite{Lyons:1988rp, Nisius:2020jmf}.
Not included in this combination are recent measurements of \mt performed with a partial 13\TeV data set~\cite{CMS:TOP-17-007, CMS:TOP-17-008, CMS:TOP-17-001, CMS:TOP-19-005, CMS:TOP-19-009, ATLAS:TOPQ-2017-17, CMS:TOP-20-008, CMS:TOP-23-003}, for which the correlations have not yet been studied in detail. These measurements include new analysis techniques and the most precise measurement to date, made by CMS, with an uncertainty of 0.37\GeV~\cite{CMS:TOP-20-008}.

The final state of events containing top quarks is determined by the decay of the \PW bosons produced in the top quark decays.
In the top quark pair (\ttbar) production mode, ATLAS and CMS have made measurements in the dilepton ($\ttbar\to\Pell^+\PGn\PQb\Pell^-\PAGn\PAQb$), lepton+jets ($\ttbar\to\Pell^{\pm}\PGn\PQb\qqbar^\prime\PAQb$), and all-jets ($\ttbar\to\qqbar^\prime\PQb\qqbar^\prime\PAQb$) channels.
In addition, CMS has performed a measurement using single top quark ($\PQt\to\Pell^+\PGn\PQb$, $\PAQt\to\Pell^-\PAGn\PAQb$) events.

In the dilepton channel, ATLAS uses the average invariant mass of the two lepton and \PQb-tagged jet pairs as the observable sensitive to \mt~\cite{ATLAS:TOPQ-2013-02, ATLAS:TOPQ-2016-03}, where a \PQb-tagged jet is any reconstructed jet identified as being likely to originate from a \PQb quark. At $\sqrt{s}=7\TeV$, CMS uses a kinematic reconstruction with the analytical matrix weighting technique~\cite{CMS:TOP-11-016} and at $\sqrt{s}=8\TeV$ CMS performs a fit to two dedicated observables~\cite{CMS:TOP-15-008} to simultaneously extract \mt and the global jet energy scale (JES).
In the lepton+jets channel~\cite{ATLAS:TOPQ-2013-02, ATLAS:TOPQ-2017-03,CMS:TOP-11-015, CMS:TOP-14-022}, both experiments perform a kinematic fit on an event-by-event basis to reconstruct the top quark mass and the invariant mass of the hadronically decaying \PW boson. The latter is used to constrain the global JES.
In addition, ATLAS fits a scale factor for the relative JES between jets originating from \PQb quarks (\PQb quark jets) and light quark / gluon jets~\cite{ATLAS:TOPQ-2013-02, ATLAS:TOPQ-2017-03}.
For the all-jets channel, ATLAS uses the ratio of the reconstructed top quark mass to the
reconstructed \PW boson mass~\cite{ATLAS:TOPQ-2013-03, ATLAS:TOPQ-2015-03} to extract \mt,
while CMS fits the reconstructed top quark mass~\cite{CMS:TOP-11-017} directly,
and at 8\TeV exploits the larger data sample to constrain the global JES using the reconstructed \PW boson mass~\cite{CMS:TOP-14-022}.

The CMS single top quark analysis fits the invariant mass of the lepton, neutrino, and \PQb-tagged jet~\cite{CMS:TOP-15-001} to extract \mt.
Two additional CMS measurements~\cite{CMS:TOP-12-030,CMS:TOP-15-014} use observables built only from leptons and charged-particle tracks, resulting in \mt measurements with low sensitivity to the JES uncertainties.
The \PJGy analysis uses the invariant mass of the lepton and the two muons from the \PJGy meson decay~\cite{CMS:TOP-15-014}.
The secondary vertex analysis uses the invariant mass of the lepton and the charged particles from a displaced secondary vertex~\cite{CMS:TOP-12-030}.
Both measurements use \ttbar events from the dilepton and lepton+jets decay modes.

All \mt measurements are calibrated using Monte Carlo (MC) simulation.
Matrix element (ME) calculations are performed at fixed order in quantum chromodynamics (QCD) and interfaced to a parton shower (PS) algorithm that provides resummation of soft and collinear QCD radiation and a hadronization model that simulates the nonperturbative formation of hadrons.
The \POWHEG~\cite{Nason:2004rx, Frixione:2007vw, Alioli:2010xd} generator at next-to-leading-order (NLO) in the strong coupling constant is interfaced with \PYTHIA{6}~\cite{Sjostrand:2006za} to simulate \ttbar production in the ATLAS measurements.
The CMS measurements use the \MADGRAPH{5}~\cite{Alwall:2011uj} generator, which includes leading-order (LO) terms for \ttbar production with up to three additional partons, also interfaced with \PYTHIA{6}.
The top quark mass is a renormalization-scheme-dependent parameter in perturbative quantum field theory. The precise identification of the \mt parameter in MC simulations within a field-theoretic mass scheme is the subject of theoretical studies~\cite{Hoang:2018zrp, Azzi:2019yne, Hoang:2020iah, Dehnadi:2023msm}.

The BLUE method defines the estimator $\mt = \sum_i w^i \mt^i$ for the input measurements $\mt^i$.
The weights $w^i$ are determined by minimizing the uncertainty in \mt, where the covariance between each pair of measurements is the crucial input.
The individual analyses $i$ are defined to be orthogonal, such that each measurement is statistically uncorrelated with every other measurement.
The exception is the CMS secondary vertex analysis~\cite{CMS:TOP-12-030}, which overlaps statistically with the dilepton and lepton+jets measurements~\cite{CMS:TOP-14-022, CMS:TOP-15-008}.
Given the different nature of the observable in the secondary vertex analysis, the analyses are assumed to be uncorrelated.
Taking the maximal statistical correlation allowed by the overlap produces no significant impact on the combination.

The measurements are affected by similar systematic uncertainties, and the assessment of their correlation is central to the combination.
As the treatment of systematic uncertainties differs between ATLAS and CMS, for each measurement they are mapped onto 25 categories that group together similar sources of uncertainties.
Uncertainty categories can influence \mt in opposite directions for different measurements, as seen in the ATLAS combinations~\cite{ATLAS:TOPQ-2013-02, ATLAS:TOPQ-2017-03}, and this effect is included via negative correlations.
The correlations between pairs of measurements from a single experiment for each category are evaluated by summing the covariance matrices of all the input uncertainty sources, mainly using the correlation assumptions discussed in Refs.~\cite{ATLAS:TOPQ-2017-03, CMS:TOP-14-022}.
Differences relative to Refs.~\cite{ATLAS:TOPQ-2017-03, CMS:TOP-14-022} are discussed in \refsupplemental.
Each input uncertainty source is included irrespective of whether it is statistically significant.

The correlation strength $\rho$ between ATLAS and CMS for each uncertainty category is assessed based on the similarities of the underlying models and methods, and of the estimates used.
Three different cases are identified, with corresponding assumed correlation strengths: $\rho=0.85$ (strongly correlated), $\rho=0.5$ (partially correlated), and $\rho=0$ (uncorrelated).
No category was identified to have $\rho = 1$, which reflects the many differences between the two experiments.
The correlation coefficient between an ATLAS and CMS measurement for each  category is the product of the respective correlation strength and the signs of the impacts of that category on each measurement.
In this way, for a given pair of measurements, categories that impact \mt in the same (opposite) direction have a positive (negative) correlation.
For categories composed of multiple uncertainty sources (\eg, \PQb tagging in ATLAS), the sign of the combined impact is not determined.
In this case, the sign of the combined impact is assumed to be positive and it was checked that taking the alternative assumption of a negative sign does not significantly impact the result, with the largest change in the central value (uncertainty) being 41 (7)\MeV.
In calculating the final covariance matrix, it is assumed that each category is uncorrelated to the others.

Table~\ref{tab:LHC_correl} displays the correlation strengths between ATLAS and CMS for each systematic uncertainty category, and \refsupplemental provides tables with the uncertainties for all 15 measurements. The corresponding correlation coefficients are available in \mbox{HEPData}~\cite{Maguire:2017ypu, hepdata}.
The subsequent paragraphs outline the categorization of systematic uncertainties and their corresponding correlation assessments.

\begin{table}[!htb]
\centering
\topcaption{%
    Correlation strengths $\rho$ of the systematic uncertainty categories between ATLAS and CMS, as used in the combination.
    The categories are defined in the text.
    Categories indicated with the symbol \NA\ in the second column correspond to uncertainties specific to a single experiment.
    The third column shows the range of $\rho$ scanned for stability checks.
    The changes in the combination's central value \mt and uncertainty \sigmt corresponding to each correlation variation are shown in the last two columns.}
\label{tab:LHC_correl}
\renewcommand\arraystretch{1.1}
\begin{scotch}{lcccc}
    \multirow{2}{*}{\shortstack{Uncertainty category}} & \multirow{2}{*}{$\rho$} & \multirow{2}{*}{Scan range} & $\Delta\mt/2$ & $\Delta\sigmt/2$ \\[-2pt]
    & & & [{\MeVns}] & [{\MeVns}] \\
    \hline
    \LHCJES1 & 0 & \NA & \NA & \NA \\
    \LHCJES2 & 0 & $[-0.25, +0.25]$ & 8 & 7 \\
    \LHCJES3 & 0.5 & $[+0.25, +0.75]$ & 1 & $<$1 \\
    \LHCbJES & 0.85 & $[+0.5, +1]$ & 26 & 5 \\
    \LHCgJES & 0.85 & $[+0.5, +1]$ & 2 & $<$1 \\
    \LHClJES & 0 & $[-0.25, +0.25]$ & 1 & $<$1 \\
    \CMSJES1 & \NA & \NA & \NA & \NA \\
    JER & 0 & $[-0.25, +0.25]$ & 5 & 1 \\
    Leptons & 0 & $[-0.25, +0.25]$ & 2 & 2 \\
    \PQb tagging & 0.5 & $[+0.25, +0.75]$ & 1 & 1 \\
    \ptmiss & 0 & $[-0.25, +0.25]$ & $<$1 & $<$1 \\
    Pileup & 0.85 & $[+0.5, +1]$ & 2 & $<$1 \\
    Trigger & 0 & $[-0.25, +0.25]$ & $<$1 & $<$1 \\ [\cmsTabSkip]
    ME generator & 0.5 & $[+0.25, +0.75]$ & $<$1 & 4 \\
    QCD radiation & 0.5 & $[+0.25, +0.75]$ & 7 & 1 \\
    Hadronization & 0.5 & $[+0.25, +0.75]$ & 1 & $<$1 \\
    CMS \PQb hadron \BR & \NA & \NA & \NA & \NA \\
    Color reconnection & 0.5 & $[+0.25, +0.75]$ & 3 & 1 \\
    Underlying event & 0.5 & $[+0.25, +0.75]$ & 1 & $<$1 \\
    PDF & 0.85 & $[+0.5, +1]$ & 1 & $<$1 \\
    CMS top quark \pt & \NA & \NA & \NA & \NA \\ [\cmsTabSkip]
    Background (data) & 0 & $[-0.25, +0.25]$ & 8 & 2 \\
    Background (MC) & 0.85 & $[+0.5, +1]$ & 2 & $<$1 \\ [\cmsTabSkip]
    Method & 0 & \NA & \NA & \NA \\
    Other & 0 & \NA & \NA & \NA \\ [\cmsTabSkip]
\end{scotch}
\end{table}

The JES uncertainty is important in many \mt measurements.
Six categories are used to describe the uncertainties associated with the calibration of the JES that are in common between the experiments~\cite{CMS:JME-14-003, CMS:JME-15-001, ATLAS:PERF-2014-02, CMS:JME-13-004}.
The category \LHCJES1 includes contributions from the limited size of the data samples used to derive the JES corrections and contributions due to pileup and its time-dependent variation.
For ATLAS (7 \TeV only), it also includes an uncertainty term from the effects of close-by jet activity.
This category is uncorrelated between ATLAS and CMS measurements.
The category \LHCJES2 corresponds to the uncertainties from the absolute JES determined using $\PGg/\PZ$+jets events that are not included in \LHCJES1.
There are significant differences between the ATLAS and CMS approaches~\cite{CMS:JME-14-003, CMS:JME-15-001}, including differences in the jet radius, treatment of muons in jets, and methods to correct for additional radiation.
Hence, this category is treated as uncorrelated.
The category \LHCJES3 corresponds to the modeling uncertainty in the relative $\eta$~intercalibration~\cite{ATLAS:PERF-2014-02, CMS:JME-13-004}.
Both experiments use dijet events for this calibration, and the modeling uncertainty originates from the use of different generators to predict the radiation patterns in these events.
As similar but not identical generators and techniques are used in both experiments, \LHCJES3 is treated as partially correlated.

The remaining JES categories correspond to the flavor-dependent calibration uncertainties.
The category \LHCbJES corresponds to the jet energy response uncertainty for \PQb quark jets.
The category \LHCgJES corresponds to the uncertainty in the jet response of gluon jets for CMS and the uncertainty in the difference of the jet response of gluons to light-quark (\PQu, \PQd, \PQs, \PQc) jets for ATLAS.
In both cases, MC comparisons determine the flavor-dependent effects, hence a strong correlation is used for the \LHCbJES and \Pg-JES components.
The category \LHClJES includes the combined CMS uncertainty in the jet response of light-quark jets and the ATLAS uncertainty for the flavor composition of jets in \ttbar events.
As these uncertainty sources are different, the \LHClJES component is treated as uncorrelated.
One additional flavor uncertainty category \CMSJES1 is included for the CMS 7\TeV measurements, corresponding to the full envelope of the response dependencies for gluons and all quark flavors.

The jet energy resolution (JER) uncertainty affects all measurements, and one category is used for the corresponding uncertainties.
ATLAS and CMS both measure the JER using data~\cite{ATLAS:PERF-2014-02,CMS:JME-13-004}, hence this category is treated as uncorrelated.

The energy scale, efficiency, and resolution of leptons affect the \mt measurements, and one category is used for the corresponding uncertainties.
ATLAS and CMS both calibrate the lepton energy scales, resolutions, and efficiencies using resonances that decay into dilepton pairs.
Since the calibration samples are independent between the two experiments, and
detector technologies and reconstruction algorithms are different, this category is treated as uncorrelated.

The selection criteria for many top quark measurements make use of \PQb tagging.
The uncertainty in the efficiency and rejection rate of these algorithms can impact the \mt measurements, and one category is used for the corresponding uncertainties.
Both collaborations use dijet events to calibrate the \PQb-tagging efficiency, employing equivalent methods~\cite{ATLAS:PERF-2012-04,CMS:BTV-12-001} that depend on similar simulation setups.
As the ATLAS \PQb jet calibration (unlike the CMS one) also uses \ttbar events~\cite{ATLAS:PERF-2012-04}, this category is assessed as partially, rather than strongly, correlated.

The missing transverse momentum (\ptmiss) is estimated in the two experiments with different algorithms.
Thus, the uncertainty in the \ptmiss scale originating from energy deposits not included in the reconstruction of jets or leptons is treated as uncorrelated.

The high instantaneous luminosity of the LHC results in multiple interactions in each bunch crossing (pileup).
As the modeling of pileup relies on simulation, the correlation between ATLAS and CMS is assessed to be strong.
While for other categories, the correlation strength is independent of the data set,
the pileup category has zero correlation between analyses performed at 7 and 8\TeV due to the different pileup conditions in the two data sets.

The uncertainty in the efficiency of the triggers used to select events typically have a small impact on the measurements.
As the triggers are calibrated in independent data sets, the uncertainty is treated as uncorrelated between ATLAS and CMS.

The \mt measurements rely on MC simulation of \ttbar events to relate the reconstructed observables to \mt.
The corresponding modeling uncertainties are encompassed in seven uncertainty categories.
The category ME generator includes uncertainties originating from the choice of the ME generator.
ATLAS assesses this uncertainty by comparing the results obtained using an \MCATNLO~\cite{Frixione:2002ik, Frixione:2003ei} sample with the \POWHEG sample.
CMS assesses this uncertainty by comparing the results obtained using a \POWHEG sample with the \MADGRAPH sample.
As the experiments employ different nominal MC models, the category is treated as partially correlated.
The category QCD radiation includes uncertainty sources for the modeling of QCD radiation in \ttbar events.
For the ATLAS measurements, samples with parameter variations of the initial- and final-state radiation in \PYTHIA, and the \hdamp parameter in \POWHEG
(which controls ME/PS matching and effectively regulates high-\pt QCD radiation) are used to evaluate these uncertainties.
For the CMS measurements, samples with variations of the factorization, renormalization, and matching scales are used.
Similarly to the ME category, the QCD radiation category is treated as partially correlated between the two experiments.

In the ATLAS analyses, the uncertainty originating from the hadronization model is evaluated by using an alternative PS and hadronization generator (\POWHEG{+}\HERWIG{6}~\cite{Corcella:2000bw}).
CMS addresses similar uncertainties by separately varying the \PQb quark fragmentation function and the semileptonic branching ratios (CMS \PQb hadron \BR).
As the ATLAS approach changes many aspects of the simulation that are not changed in the two CMS uncertainty sources and the \PYTHIA settings in the two experiments are not the same, there is no clear mapping and correlation for these sources.
Nevertheless, some degree of correlation is expected, hence
the ATLAS hadronization uncertainty is grouped with the CMS uncertainty from the fragmentation model in the category hadronization and this category is assumed to be partially correlated between the experiments.
The CMS \PQb hadron \BR uncertainty source is treated as uncorrelated with the ATLAS uncertainties.
It was verified that the alternative treatment of correlating the ATLAS hadronization uncertainty with the CMS \PQb hadron \BR uncertainty had no
significant impact on the result.

The uncertainties associated with color reconnection and the underlying event tunes are included in separate categories.
The experiments use different \PYTHIA settings for the nominal simulation, and these uncertainty categories are taken to be partially correlated.
The uncertainty in the parton distribution functions (PDFs) is driven by the input data used in the PDF extractions, and hence this category is taken as strongly correlated between ATLAS and CMS.
The CMS analyses account for an uncertainty in the modeling of the top quark \pt distribution, represented by a separate category, while for the ATLAS analyses, the alternative MC sample used to evaluate the hadronization uncertainty covers the disagreement between data and simulation~\cite{ATLAS:TOPQ-2015-06}, and no additional uncertainty is evaluated.

The analyses typically have small contributions from background processes, and background uncertainties have only a small impact on the measurements.
Uncertainties in backgrounds estimated from data control samples are included in the category Background (data), treated as uncorrelated between the experiments.
Both ATLAS and CMS rely on MC simulation for several backgrounds.
The uncertainties in these are included in the category Background (MC), assumed to be strongly correlated.

Every analysis ensures that the \mt fit is unbiased.
This is done using simulated samples generated with different \mt values.
The limited sample size introduces a systematic uncertainty (Method) that is statistical and hence uncorrelated between measurements.

A few systematic uncertainties affect only a limited number of analyses (see \refsupplemental).
These uncertainty sources are in the category Other, which is uncorrelated between ATLAS and CMS.

The measurements from each experiment are separately combined, with the ATLAS combination giving $\mt = 172.71 \pm 0.25 \stat \pm 0.41 \syst \GeV$ and the CMS combination giving $\mt = 172.52 \pm 0.14 \stat \pm 0.39 \syst \GeV$.
The ATLAS combination is very similar to, and supersedes, the result in Ref.~\cite{ATLAS:TOPQ-2017-03}, with the slight difference originating from changes in the correlation assumptions
that are discussed in \refsupplemental.
The CMS measurement is improved compared to the previous combination~\cite{CMS:TOP-14-022} and supersedes that result.
The improvement originates from including a more precise dilepton measurement at 8\TeV together with the single top, secondary vertex, and \PJGy meson measurements,
and from including the effect of anticorrelations of the systematic uncertainties between the input measurements.
It was verified that performing the combinations with a likelihood-based approach~\cite{Kieseler:2017kxl} gives identical results.

\begin{figure}[!tb]
\centering
\ifthenelse{\boolean{cms@external}}{%
    \includegraphics[width=\linewidth]{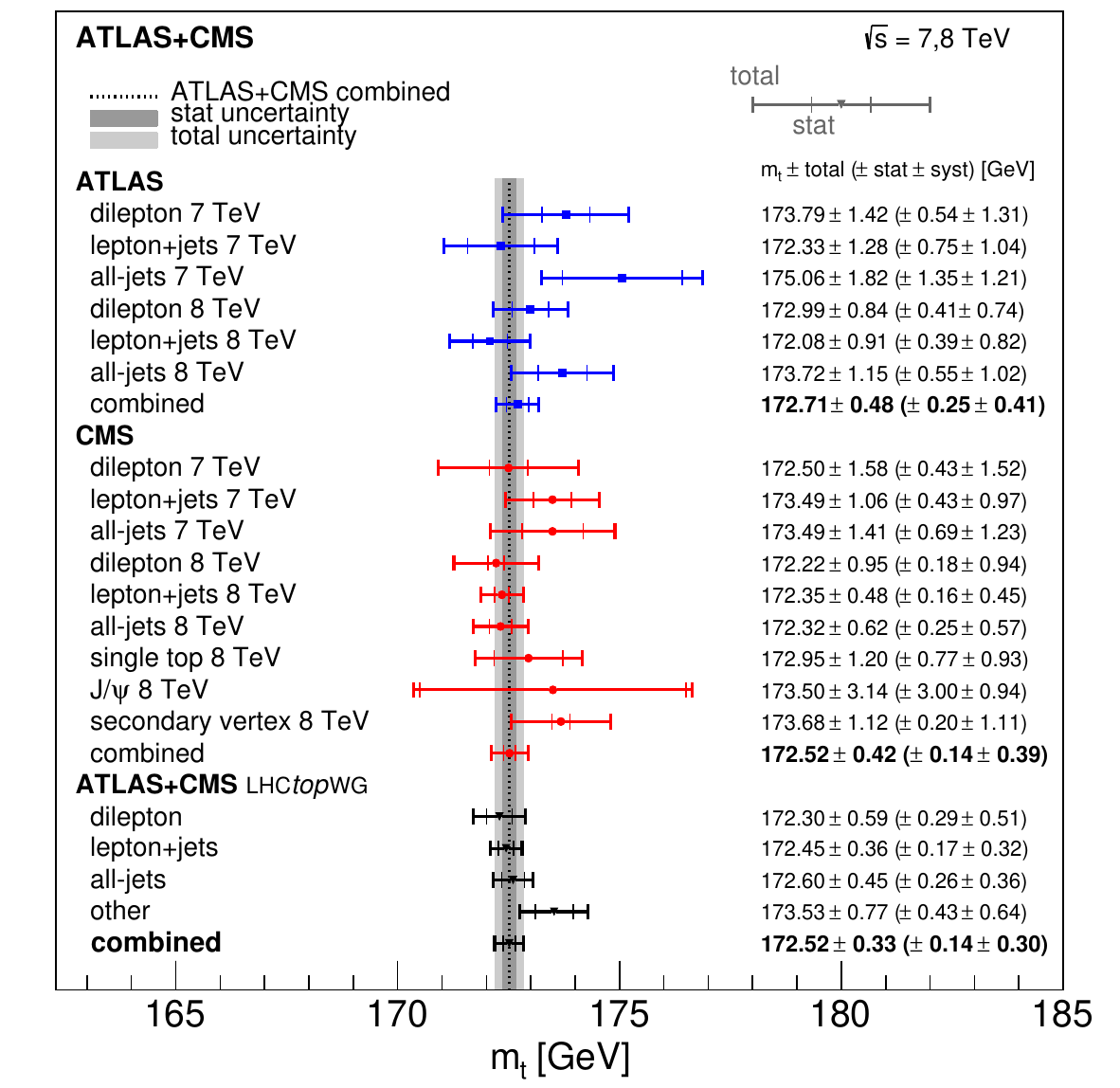}%
}{%
    \includegraphics[width=0.7\textwidth]{Figure_001.pdf}%
}
\caption{%
    Comparison of the individual \mt measurements and the result of the \mt combination. Also shown are the separate combinations of each experiment
    and the result of the simultaneous combination for the different decay channels, where the ``other'' category covers the single top, \JPsi, and secondary vertex measurements.
}
\label{fig:LHC_mt_summary}
\end{figure}

The combination of all 15 input measurements gives
\begin{linenomath}\begin{equation*}
    \mt = 172.52 \pm 0.14 \stat \pm 0.30 \syst \GeV,
\end{equation*}\end{linenomath}
which is compared with the input measurements in Fig.~\ref{fig:LHC_mt_summary}. The LHC combination has the same statistical uncertainty as the CMS combination.
This is because the figure of merit in BLUE is the total uncertainty, and the statistical component is a consequence of the optimized weights in the combination.
The difference in statistical uncertainty between ATLAS and CMS reflects different analysis choices, as explained in \refsupplemental.

The combination achieves an improvement in the total \mt uncertainty of 31\% relative to the most precise input measurement.
The measurements with the largest weight in the combination are the CMS 8\TeV lepton+jets (0.34), dilepton (0.12), and all-jets (0.12) results, and the ATLAS 8\TeV lepton+jets (0.17) and dilepton (0.16) measurements.
The hierarchy of the weights originates from the uncertainty of each measurement, as well as the correlation between measurements. For example, the ATLAS 8\TeV lepton+jets measurement has a higher weight than the corresponding dilepton measurement, despite having a larger uncertainty. This is because of the smaller correlation with the precise CMS 8\TeV lepton+jets measurement.
The combination shows good compatibility between the measurements, with $\chisq=7.5$ and a corresponding $p$-value of 91\%.
The LHC combination is much closer to the CMS combination than the ATLAS one because the relative weights of the measurements with slightly lower measured \mt are higher in the LHC combination than in the per-experiment combinations.
All weights and the individual pulls can be found in \refsupplemental, along with a combination where all 15 measurements are used to extract separate \mt values for ATLAS and CMS.

\begin{table}[!tb]
\centering
\topcaption{%
    Uncertainties on the \mt values extracted in the LHC, ATLAS, and CMS combinations
    arising from the same categories as listed in Table~\ref{tab:LHC_correl}, sorted
    in order of decreasing value of the combined LHC uncertainty.
}
\label{tab:MethodAimpacts}
\renewcommand\arraystretch{1.1}
\begin{scotch}{lccc}
\multirow{2}{*}{Uncertainty category} & \multicolumn{3}{c}{Uncertainty impact [{\GeVns}]} \\[-1pt]
& ~LHC~ & ATLAS & ~CMS~ \\ \hline
               \LHCbJES    &   0.18    &   0.17    &   0.25 \\
           \PQb tagging    &   0.09    &   0.16    &   0.03 \\
           ME generator    &   0.08    &   0.13    &   0.14 \\
               \LHCJES1    &   0.08    &   0.18    &   0.06 \\
               \LHCJES2    &   0.08    &   0.11    &   0.10 \\
                 Method    &   0.07    &   0.06    &   0.09 \\
    CMS \PQb hadron \BR    &   0.07    &   \NA     &   0.12 \\
          QCD radiation    &   0.06    &   0.07    &   0.10 \\
                Leptons    &   0.05    &   0.08    &   0.07 \\
                    JER    &   0.05    &   0.09    &   0.02 \\
      CMS top quark \pt    &   0.05    &   \NA     &   0.07 \\
      Background (data)    &   0.05    &   0.04    &   0.06 \\
     Color reconnection    &   0.04    &   0.08    &   0.03 \\
       Underlying event    &   0.04    &   0.03    &   0.05 \\
               \LHCgJES    &   0.03    &   0.02    &   0.04 \\
        Background (MC)    &   0.03    &   0.07    &   0.01 \\
                  Other    &   0.03    &   0.06    &   0.01 \\
               \LHClJES    &   0.03    &   0.01    &   0.05 \\
               \CMSJES1    &   0.03    &   \NA     &   0.04 \\
                 Pileup    &   0.03    &   0.07    &   0.03 \\
               \LHCJES3    &   0.02    &   0.07    &   0.01 \\
          Hadronization    &   0.02    &   0.01    &   0.01 \\
                \ptmiss    &   0.02    &   0.04    &   0.01 \\
                    PDF    &   0.02    &   0.06    &   \makebox[0pt][r]{$<$}0.01 \\
                Trigger    &   0.01    &   0.01    &   0.01 \\[\cmsTabSkip]
       Total systematic    &   0.30    &   0.41    &   0.39 \\
            Statistical    &   0.14    &   0.25    &   0.14 \\[\cmsTabSkip]
                  Total    &   0.33    &   0.48    &   0.42 \\
\end{scotch}
\end{table}

Table~\ref{tab:MethodAimpacts} shows the breakdown of the systematic uncertainty in the combined measurement and the individual ATLAS and CMS combinations.
The largest systematic uncertainties are seen to originate from JES, \PQb tagging, and \ttbar modeling.
The stability of the measurement against the correlation assumptions is checked by varying the correlation strengths for each uncertainty category as shown in Table~\ref{tab:LHC_correl}.
The ranges reflect the extent of the understanding of the correlations.
No variation is performed for categories where there is no ambiguity in the correlation assumption.
Table~\ref{tab:LHC_correl} shows the variation in the total uncertainty and central value of the combination under those changes.
Both the central value and uncertainty are observed to vary linearly under the variations and the changes are small ($<$30\MeV) compared to the uncertainty in \mt.
The largest change in central value is seen for \LHCbJES, which is the leading correlated uncertainty source in the combination.

The consistency of the result and the measurements from the different decay channels have been checked by performing the combination with a separate \mt parameter for each \ttbar decay channel.
The results are also shown in Fig.~\ref{fig:LHC_mt_summary}, and the \mt values are found to be consistent.

The impact of the limited statistical precision of the estimates of the systematic uncertainties is evaluated by performing pseudo-experiments where the systematic uncertainties of the measurements are varied according to their uncertainties and the combination procedure is repeated. In this procedure, changes in the sign of the impacts of systematic uncertainties are propagated to the signs of the corresponding correlations.
The root-mean-square of the measured \mt (\sigmt) is found to be 63 (19)\MeV, demonstrating the stability of the combination.

The understanding of top quark production and decay has continued to evolve since the publication of the measurements used in this combination.
Developments in the simulations include improved modeling of off-shell effects~\cite{Jezo:2016ujg}, reduced uncertainties in additional QCD radiation~\cite{ATLAS:PHYS-PUB-2018-009, CMS:TOP-16-021}, new models of color reconnection~\cite{Argyropoulos:2014zoa, Christiansen:2015yqa}, MC simulations at next-to-NLO precision in QCD~\cite{Mazzitelli:2021mmm}, and investigations into the radiation patterns in the top quark decay~\cite{Brooks:2019xso}.
Advancements in the modeling, which may either increase or decrease the mass uncertainty, and improvements in analysis techniques~\cite{CMS:TOP-17-001,CMS:TOP-20-008} are being incorporated into analyses performed at $\sqrt{s}=13\TeV$, but this is not possible for the analyses used in this combination.
A cross-check, detailed in \refsupplemental, was performed to verify that potential modeling uncertainties in the recoil in the top quark decay~\cite{Brooks:2019xso}
do not significantly affect the combination.

In summary, a combination of top quark mass measurements by the ATLAS and CMS experiments at the CERN LHC in proton-proton collisions at $\sqrt{s}=7$ and 8\TeV has been performed.
The combination yields $\mt = 172.52 \pm 0.33\GeV$, which is the most precise result to date.

\begin{acknowledgments}
CMS congratulates our colleagues in the CERN accelerator departments for the excellent performance of the LHC and thank the technical and administrative staffs at CERN and at other CMS institutes for their contributions to the success of the CMS effort. In addition, we gratefully acknowledge the computing centres and personnel of the Worldwide LHC Computing Grid and other centres for delivering so effectively the computing infrastructure essential to our analyses. Finally, we acknowledge the enduring support for the construction and operation of the LHC, the CMS detector, and the supporting computing infrastructure provided by the following funding agencies: SC (Armenia), BMBWF and FWF (Austria); FNRS and FWO (Belgium); CNPq, CAPES, FAPERJ, FAPERGS, and FAPESP (Brazil); MES and BNSF (Bulgaria); CERN; CAS, MoST, and NSFC (China); MINCIENCIAS (Colombia); MSES and CSF (Croatia); RIF (Cyprus); SENESCYT (Ecuador); ERC PRG, RVTT3 and TK202 (Estonia); Academy of Finland, MEC, and HIP (Finland); CEA and CNRS/IN2P3 (France); SRNSF (Georgia); BMBF, DFG, and HGF (Germany); GSRI (Greece); NKFIH (Hungary); DAE and DST (India); IPM (Iran); SFI (Ireland); INFN (Italy); MSIP and NRF (Republic of Korea); MES (Latvia); LAS (Lithuania); MOE and UM (Malaysia); BUAP, CINVESTAV, CONACYT, LNS, SEP, and UASLP-FAI (Mexico); MOS (Montenegro); MBIE (New Zealand); PAEC (Pakistan); MES and NSC (Poland); FCT (Portugal);  MESTD (Serbia); MCIN/AEI and PCTI (Spain); MOSTR (Sri Lanka); Swiss Funding Agencies (Switzerland); MST (Taipei); MHESI and NSTDA (Thailand); TUBITAK and TENMAK (Turkey); NASU (Ukraine); STFC (United Kingdom); DOE and NSF (USA).

ATLAS thanks CERN for the very successful operation of the LHC and its injectors, as well as the support staff at CERN and at our institutions worldwide without whom ATLAS could not be operated efficiently.
 
The crucial computing support from all WLCG partners is acknowledged gratefully, in particular from CERN, the ATLAS Tier-1 facilities at TRIUMF/SFU (Canada), NDGF (Denmark, Norway, Sweden), CC-IN2P3 (France), KIT/GridKA (Germany), INFN-CNAF (Italy), NL-T1 (Netherlands), PIC (Spain), RAL (UK) and BNL (USA), the Tier-2 facilities worldwide and large non-WLCG resource providers. Major contributors of computing resources are listed in Ref.~\cite{ATL-SOFT-PUB-2023-001}.
 
We gratefully acknowledge the support of ANPCyT, Argentina; YerPhI, Armenia; ARC, Australia; BMWFW and FWF, Austria; ANAS, Azerbaijan; CNPq and FAPESP, Brazil; NSERC, NRC and CFI, Canada; CERN; ANID, Chile; CAS, MOST and NSFC, China; Minciencias, Colombia; MEYS CR, Czech Republic; DNRF and DNSRC, Denmark; IN2P3-CNRS and CEA-DRF/IRFU, France; SRNSFG, Georgia; BMBF, HGF and MPG, Germany; GSRI, Greece; RGC and Hong Kong SAR, China; ISF and Benoziyo Center, Israel; INFN, Italy; MEXT and JSPS, Japan; CNRST, Morocco; NWO, Netherlands; RCN, Norway; MEiN, Poland; FCT, Portugal; MNE/IFA, Romania; MESTD, Serbia; MSSR, Slovakia; ARRS and MIZ\v{S}, Slovenia; DSI/NRF, South Africa; MICINN, Spain; SRC and Wallenberg Foundation, Sweden; SERI, SNSF and Cantons of Bern and Geneva, Switzerland; MOST, Taipei; TENMAK, T\"urkiye; STFC, United Kingdom; DOE and NSF, United States of America.
 
Individual groups and members have received support from BCKDF, CANARIE, CRC and DRAC, Canada; CERN-CZ, PRIMUS 21/SCI/017 and UNCE SCI/013, Czech Republic; COST, ERC, ERDF, Horizon 2020, ICSC-NextGenerationEU and Marie Sk{\l}odowska-Curie Actions, European Union; Investissements d'Avenir Labex, Investissements d'Avenir Idex and ANR, France; DFG and AvH Foundation, Germany; Herakleitos, Thales and Aristeia programmes co-financed by EU-ESF and the Greek NSRF, Greece; BSF-NSF and MINERVA, Israel; Norwegian Financial Mechanism 2014-2021, Norway; NCN and NAWA, Poland; La Caixa Banking Foundation, CERCA Programme Generalitat de Catalunya and PROMETEO and GenT Programmes Generalitat Valenciana, Spain; G\"{o}ran Gustafssons Stiftelse, Sweden; The Royal Society and Leverhulme Trust, United Kingdom.

\begin{sloppypar}
In addition, individual members wish to acknowledge support from Chile: Agencia Nacional de Investigaci\'on y Desarrollo (FONDECYT 1190886, FONDECYT 1210400, FONDECYT 1230812, FONDECYT 1230987); China: National Natural Science Foundation of China (NSFC - 12175119, NSFC 12275265, NSFC-12075060); Czech Republic: PRIMUS Research Programme (PRIMUS/21/SCI/017); EU: H2020 European Research Council (ERC - 101002463); European Union: European Research Council (ERC - 948254), Horizon 2020 Framework Programme (MUCCA - CHIST-ERA-19-XAI-00), European Union, Future Artificial Intelligence Research (FAIR-NextGenerationEU PE00000013), Italian Center for High Performance Computing, Big Data and Quantum Computing (ICSC, NextGenerationEU), Marie Sklodowska-Curie Actions (EU H2020 MSC IF GRANT NO 101033496); France: Agence Nationale de la Recherche (ANR-20-CE31-0013, ANR-21-CE31-0013, ANR-21-CE31-0022), Investissements d'Avenir Idex (ANR-11-LABX-0012), Investissements d'Avenir Labex (ANR-11-LABX-0012); Germany: Baden-Württemberg Stiftung (BW Stiftung-Postdoc Eliteprogramme), Deutsche Forschungsgemeinschaft (DFG - 469666862, DFG - CR 312/5-1); Italy: Istituto Nazionale di Fisica Nucleare (FELLINI G.A. n. 754496, ICSC, NextGenerationEU); Japan: Japan Society for the Promotion of Science (JSPS KAKENHI JP21H05085, JSPS KAKENHI JP22H01227, JSPS KAKENHI JP22H04944, JSPS KAKENHI JP22KK0227); Netherlands: Netherlands Organisation for Scientific Research (NWO Veni 2020 - VI.Veni.202.179); Norway: Research Council of Norway (RCN-314472); Poland: Polish National Agency for Academic Exchange (PPN/PPO/2020/1/00002/U/00001), Polish National Science Centre (NCN 2021/42/E/ST2/00350, NCN OPUS nr 2022/47/B/ST2/03059, NCN UMO-2019/34/E/ST2/00393, UMO-2020/37/B/ST2/01043, UMO-2021/40/C/ST2/00187); Slovenia: Slovenian Research Agency (ARIS grant J1-3010); Spain: BBVA Foundation (LEO22-1-603), Generalitat Valenciana (Artemisa, FEDER, IDIFEDER/2018/048), La Caixa Banking Foundation (LCF/BQ/PI20/11760025), Ministry of Science and Innovation (MCIN \& NextGenEU PCI2022-135018-2, MICIN \& FEDER PID2021-125273NB, RYC2019-028510-I, RYC2020-030254-I, RYC2021-031273-I, RYC2022-038164-I), PROMETEO and GenT Programmes Generalitat Valenciana (CIDEGENT/2019/023, CIDEGENT/2019/027); Sweden: Swedish Research Council (VR 2018-00482, VR 2022-03845, VR 2022-04683, VR grant 2021-03651), Knut and Alice Wallenberg Foundation (KAW 2017.0100, KAW 2018.0157, KAW 2018.0458, KAW 2019.0447); Switzerland: Swiss National Science Foundation (SNSF - PCEFP2\_194658); United Kingdom: Leverhulme Trust (Leverhulme Trust RPG-2020-004); United States of America: U.S. Department of Energy (ECA DE-AC02-76SF00515), Neubauer Family Foundation.
\end{sloppypar}
\end{acknowledgments}

\bibliography{auto_generated}

\ifthenelse{\boolean{cms@external}}{
\nocite{ATLAS:2010arf, Sjostrand:2014zea, Nisius:2014wua}
}
{
\clearpage
\numberwithin{table}{section}
\numberwithin{figure}{section}
\appendix
\section{Supplemental material\label{this:Supplemental}}

\subsection{Changes to the ATLAS correlations}

Two changes in the correlation model for the ATLAS combination are made compared to Ref.~\cite{ATLAS:TOPQ-2017-03}. The \PQb tagging algorithm and calibration is different
for the 8\TeV all-jets analysis compared to the lepton+jets and dilepton measurements. This prevents a precise correlation assessment for this source.
Reference~\cite{ATLAS:TOPQ-2017-03} assumed a correlation of $+1$ for this source between the all-jets and lepton+jets/dilepton measurements, and this is modified to 0 for this
combination to reflect the use of a different algorithm. The pileup uncertainties were assumed to be uncorrelated between all the analyses in Ref.~\cite{ATLAS:TOPQ-2017-03}.
For this combination, the pileup uncertainty correlation is assumed to be $+1$ between each analysis at the same energy, reflecting the fact each analysis shares common
modeling of the pileup. Zero correlation is assumed between the 7 and 8\TeV data sets, reflecting the significant difference in the pileup conditions. These
two changes move the ATLAS combined \mt measurement by 20\MeV and have no impact on the uncertainty at the quoted precision.

\subsection{Further details on the CMS correlations}

Several changes in the correlation model for the CMS combination are made compared to Ref.~\cite{CMS:TOP-14-022}. To be
consistent with the correlation model used in the ATLAS combination, the signs of systematic effects are accounted for,
leading to negative signs for some of the correlations between analyses. In the previous CMS
combination~\cite{CMS:TOP-14-022} all correlation signs were assumed to be positive. Another change compared to
Ref.~\cite{CMS:TOP-14-022}  is that the jet flavor uncertainties are assumed to be uncorrelated to match the
ATLAS treatment, while in the original
published analyses they were treated as fully correlated. In the updated CMS combination, the effect of this change on
the central mass value (uncertainty) is 1 (10)\MeV.
The tables in the HEPData record~\cite{hepdata} show the detailed correlations used in the combination.
Sources of uncertainty of a statistical nature or that are unique to a specific analysis are considered uncorrelated between measurements. These include Background (data), Trigger, Method, and Other.
\LHCJES1 is assumed uncorrelated between 7 and 8\TeV and correlated otherwise, as it is of statistical nature but is common to measurements at the same center-of-mass energy.
Similarly, Pileup is uncorrelated between 7 and 8\TeV. All other sources of uncertainty are considered fully correlated between all measurements.

\subsection{Uncertainties in the Other category}

The following uncertainty sources are included in the Other category:
\begin{itemize}
\item The uncertainties in single top quark modeling that impact the CMS single top quark measurement.
\item The uncertainty in the modeling of the \JPsi meson candidate mass distribution that affects the CMS \JPsi measurement.
\item The uncertainty in \PQb hadron composition that impacts the CMS secondary vertex measurement.
\item Lepton reconstruction and identification uncertainties for the CMS 8\TeV dilepton measurement. The other CMS lepton+jets and dilepton measurements assume that these uncertainties are negligible.
\item The uncertainties in the efficiency of jet reconstruction and the selections used to reject jets from pileup that affect ATLAS measurements.
\item The uncertainty originating from using fast simulation~\cite{ATLAS:2010arf} in the ATLAS 7\TeV all-jets measurement. All other measurements use full simulation.
\end{itemize}

\subsection{Cross-check on modeling of the recoil in the top quark decay}

The PS simulation used in \PYTHIA has an ambiguity in the choice of recoil particle in the case of additional gluon radiation in the top quark decay~\cite{Brooks:2019xso}.
As the shower includes matrix element corrections, this ambiguity does not affect the first gluon emission, but only the subsequent emissions.
The effect described in Ref.~\cite{Brooks:2019xso} is expected to mostly change the fraction of out-of-cone radiation of \PQb quark jets.
In the \PYTHIA{6} simulations used in the measurements that enter the combination, only one recoil scheme is available.
The \PYTHIA{8} generator~\cite{Sjostrand:2014zea} offers the possibility to change the recoil scheme.
Studies using generator-level \PYTHIA{8} samples (generated at $\sqrt{s}=13\TeV$) show
that changes in the reconstructed \mt seen when changing the recoil scheme from \verb|RecoilToColoured=On| to recoil-to-top~\cite{ATLAS:TOPQ-2017-17}
correspond to a change in the \PQb jet energy that is around 70\% of the \LHCbJES uncertainty.
As a further cross-check, the combination was repeated by adding an additional uncertainty to each analysis equal to 70\% of the \LHCbJES uncertainty.
The central value and uncertainty of the combination are observed to increase by 35 and 20\MeV respectively. These values are small compared to the total uncertainty,
indicating that the result is robust against potential additional modeling uncertainties such as this.

\subsection{Simultaneous ATLAS and CMS top quark mass combination}

The compatibility of the ATLAS and CMS results is evaluated by performing a combination using all 15 input measurements with one \mt per experiment (\mtATLAS and \mtCMS).
This combination yields $\mtATLAS = 172.72 \pm 0.25 \stat \pm 0.39 \syst \GeV$ and $\mtCMS = 172.37 \pm 0.14 \stat \pm 0.38 \syst \GeV$, with excellent agreement seen between the two experiments.
The larger statistical uncertainty in the ATLAS combination reflects the larger statistical uncertainties in the ATLAS analyses compared to the CMS analyses in the same channels,
which in turn reflect the different analysis methods and choices.
For example, the ATLAS lepton+jets measurements~\cite{ATLAS:TOPQ-2013-02, ATLAS:TOPQ-2017-03,CMS:TOP-11-015} fit a relative \PQb-to-light JES factor,
effectively trading smaller systematic uncertainty for larger statistical uncertainty.
This result of the simultaneous combination is displayed in Fig.~\ref{fig:lhc2d}, 
which compares the results of this simultaneous combination with the expectation $\mtLHC=\mtATLAS=\mtCMS$, 
the result of the default LHC combination using all 15 input measurements with a single 
extracted \mt (filled circle), 
and the results of the individual experiment combinations (red and blue lines), where
\mtATLAS and \mtCMS are extracted from separate combinations to the 6 ATLAS and 9 CMS measurements, respectively.
In the simultaneous combination of all 15 measurements, all measurements contribute to the two parameters \mtATLAS and \mtCMS, \ie, $\mtATLAS = \sum_i w_i m_i + \sum_j \lambda_j m_j$,
where $m_i$ are the ATLAS measurements and $m_j$ are the CMS measurements. The weights, shown in Table~\ref{tab:weights_subcomb_exp}, satisfy $\sum_i w_i = 1$ and $\sum_j \lambda_j=0$,
and the equivalent condition applies to the weights for \mtCMS.
The \chisq of this simultaneous combination is 7.2 (13 degrees of freedom), demonstrating no significant improvement in \chisq over the single parameter combination (7.5 for 14 degrees of freedom).
This reflects the excellent agreement between the ATLAS and CMS measurements of \mt.

\begin{figure*}[!hp]
\centering
\includegraphics[width=0.7\textwidth]{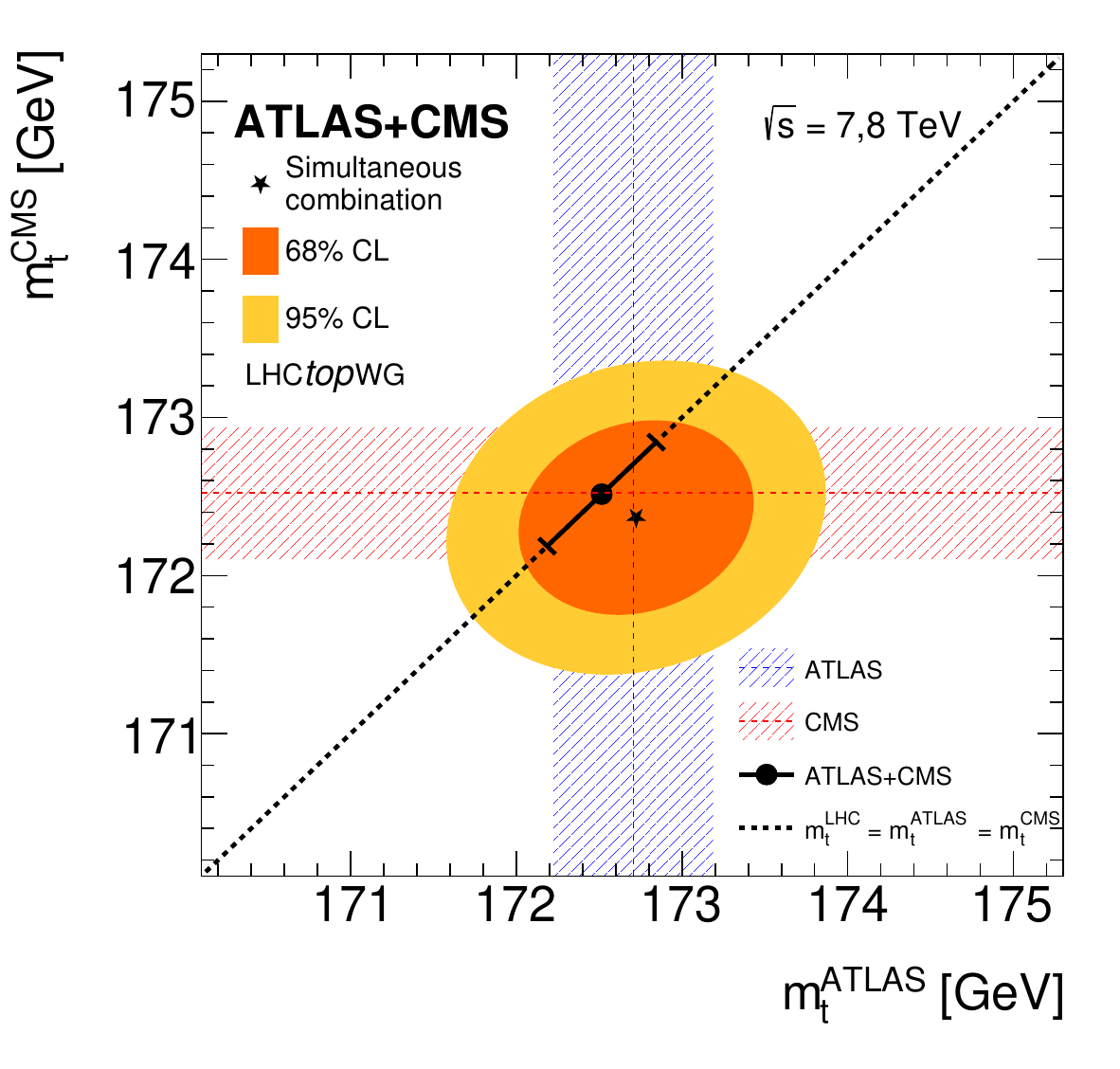}
\caption{%
   The simultaneous extraction of the \mt measured by ATLAS (\mtATLAS) and CMS (\mtCMS) from a BLUE combination of the 15 input measurements is shown by the star. The solid ellipses show the regions allowed at 68 and 95\% confidence level (\CL) by the combination and are in good agreement with the
   expectation $\mtATLAS=\mtCMS$ (shown by the black dashed line). The observed correlation between \mtATLAS and \mtCMS  is 0.15.
   The dashed blue and red lines and hashed bands show the central values and 68\% \CL intervals for the individual ATLAS and CMS combinations, which use the 6 ATLAS and 9 CMS measurements, respectively.
   In addition, the central value of the LHC combination, \mtLHC, which assumes $\mtLHC=\mtATLAS=\mtCMS$, is shown by the circular marker.
   The projection of the corresponding diagonal error bar on either axis represents the total uncertainty \mtLHC.
}
\label{fig:lhc2d}
\end{figure*}

\begin{table*}[!hp]
\centering
\topcaption{%
    BLUE weights of the simultaneous ATLAS and CMS combination for each input measurement. The input measurements are the ATLAS and CMS 7 and 8\TeV \mt measurements in the dilepton (``dil''), lepton+jets (``lj''), and all-jets (``aj'') channels, and the CMS 8\TeV \mt measurements in the single top (``t''), secondary vertex (``vtx''), and \PJGy analysis (``\PJGy''). The sum of the ATLAS weights in the CMS combined value is zero, and vice versa. The individual weights, however, are different from zero due to the correlation between the different experiments. The weights are rounded to two decimal places; when the full precision is used, the weights for each of \mtATLAS and \mtCMS sum to one.
}
\label{tab:weights_subcomb_exp}
\renewcommand\arraystretch{1.2}
\cmsTable{\begin{scotch}{lccc@{\hspace{4\tabcolsep}}ccccccc@{\hspace{4\tabcolsep}}cccccc}
& \multicolumn{6}{c}{ATLAS} & & \multicolumn{9}{c}{CMS}  \\
\cline{2-7}\cline{9-17}
& \multicolumn{3}{c}{2011 (7\TeV)} & \multicolumn{3}{c}{2012 (8\TeV)} & & \multicolumn{3}{c}{2011 (7\TeV)} &  \multicolumn{6}{c}{2012 (8\TeV)} \\
& \multicolumn{1}{c}{dil} & \multicolumn{1}{c}{lj} & \multicolumn{1}{c@{\hspace{4\tabcolsep}}}{aj} & \multicolumn{1}{c}{dil} & \multicolumn{1}{c}{lj} & \multicolumn{1}{c}{aj} & & \multicolumn{1}{c}{dil} & \multicolumn{1}{c}{lj} & \multicolumn{1}{c@{\hspace{4\tabcolsep}}}{aj} & \multicolumn{1}{c}{dil} & \multicolumn{1}{c}{lj} & \multicolumn{1}{c}{aj} & \multicolumn{1}{c}{\PQt} & \multicolumn{1}{c}{\PJGy} & \multicolumn{1}{c}{vtx} \\
\hline
\mtATLAS & $<$0.01 & $+0.16$ & $+0.04$ & $+0.33$ & $+0.36$ & $+0.11$ & & $-0.05$ & $-0.07$ & $+0.03$ & $+0.03$ & $-0.11$ & $+0.14$ & $-0.03$ & $+0.01$ & $+0.05$ \\
\mtCMS & $-0.04$ & $+0.01$ & $-0.03$ & $+0.04$ & $+0.04$ & $-0.02$ & & $-0.10$ & $+0.02$ & $+0.04$ & $+0.18$ & $+0.67$ & $+0.10$ & $-0.04$ & $+0.01$ & $+0.11$ \\
\end{scotch}}
\end{table*}

\subsection{Numerical details of the combination}

Tables~\ref{tab:data_input_ATLAS} and \ref{tab:data_input_CMS} show the ATLAS and CMS measurements together with the breakdown of their respective uncertainties, while Table~\ref{tab:pulls_weights_LHC} shows the pull and weight of each input measurement in the LHC combination, as obtained with the BLUE method.
The BLUE method can result in negative weights, as seen in Table~\ref{tab:pulls_weights_LHC}.
This typically happens for measurements with a larger uncertainty than and high correlation to a more precise measurement~\cite{Nisius:2014wua}.
Figure~\ref{fig:LHC_corr} shows the correlation between each pair of measurements used in the LHC combination.

A simultaneous combination with one \mt parameter per each decay channel is performed to check the consistency of the result (Fig.~1 in the main document).
Table~\ref{tab:weights_subcomb_ch} shows the weights for this simultaneous combination.
The ``Other'' channel includes the CMS single top, secondary vertex, and \PJGy analyses.
The combined measurement for channel $k$ is
$m^k_t = \sum_i w_i m_i + \sum_j \lambda_j m_j$, where the sum over $i$ includes all measurements of channel $k$ and the sum over $j$ includes all other measurements. The weights
satisfy $\sum_i w_i =1$ and $\sum_j \lambda_j=0$. The correlations between the measurements result in nonzero values of the individual $\lambda_j$. The \chisq of this simultaneous
combination is 5.4 (11 degrees of freedom), corresponding to a $p$-value of 91\%.
The correlations between the \mt values extracted per channel are shown in Table~\ref{tab:corr_channels}.

\begin{table*}[!hp]
\centering
\topcaption{%
    Results and systematic uncertainties of the ATLAS \mt measurements, shown separately for the 7 and 8\TeV results in the dilepton (``dil''), lepton+jets (``lj''), and all-jets (``aj'') channels, and for their combination (``comb.'').
    All values are given in \GeVns.
}
\label{tab:data_input_ATLAS}
\renewcommand\arraystretch{1.2}
\begin{scotch}{lrrr@{\hspace{4\tabcolsep}}rrr@{\hspace{4\tabcolsep}}r}
& \multicolumn{7}{c}{ATLAS} \\
\cline{2-8}
& \multicolumn{3}{c@{\hspace{4\tabcolsep}}}{2011 (7\TeV)} & \multicolumn{3}{c@{\hspace{4\tabcolsep}}}{2012 (8\TeV)} & \multirow{2}{*}{comb.} \\
& \multicolumn{1}{c}{dil} & \multicolumn{1}{c}{lj} & \multicolumn{1}{c@{\hspace{4\tabcolsep}}}{aj} & \multicolumn{1}{c}{dil} & \multicolumn{1}{c}{lj} & \multicolumn{1}{c@{\hspace{4\tabcolsep}}}{aj} & \\
\hline
\mt & 173.79  & 172.33  & 175.06  & 172.99  & 172.08  & 173.72  & 172.71 \\ [\cmsTabSkip]
\LHCJES1 & 0.54  & 0.33  & 0.38  & 0.35  & 0.28  & 0.40  & 0.18 \\
\LHCJES2 & 0.30  & 0.30  & 0.20  & 0.41  & 0.39  & 0.42  & 0.11 \\
\LHCJES3 & 0.43  & 0.07  & 0.24  & 0.08  & 0.05  & 0.12  & 0.07 \\
\LHCbJES & 0.68  & 0.06  & 0.62  & 0.30  & 0.03  & 0.34  & 0.17 \\
\LHCgJES & 0.03  & 0.28  & 0.10  & 0.02  & 0.21  & 0.05  & 0.02 \\
\LHClJES & 0.02  & 0.24  & 0.02  & 0.01  & 0.10  & 0.06  & 0.01 \\
JER & 0.19  & 0.22  & 0.01  & 0.09  & 0.20  & 0.10  & 0.09 \\
Leptons & 0.13  & 0.04  & \NA  & 0.14  & 0.16  & 0.01  & 0.08 \\
\PQb tagging & 0.07  & 0.50  & 0.16  & 0.04  & 0.38  & 0.10  & 0.16 \\
\ptmiss & 0.04  & 0.15  & 0.02  & 0.01  & 0.05  & 0.01  & 0.04 \\
Pileup & 0.01  & 0.02  & 0.02  & 0.05  & 0.15  & 0.01  & 0.07 \\
Trigger & 0.01  & \NA  & 0.01  & \NA  & 0.01  & 0.08  & 0.01 \\ [\cmsTabSkip]
ME generator & 0.26  & 0.22  & 0.30  & 0.09  & 0.16  & 0.18  & 0.13 \\
QCD radiation & 0.47  & 0.32  & 0.22  & 0.23  & 0.08  & 0.10  & 0.07 \\
Hadronization & 0.53  & 0.18  & 0.50  & 0.22  & 0.15  & 0.64  & 0.01 \\
Color reconnection & 0.14  & 0.11  & 0.22  & 0.03  & 0.19  & 0.12  & 0.08 \\
Underlying event & 0.05  & 0.15  & 0.08  & 0.10  & 0.08  & 0.12  & 0.03 \\
PDF & 0.10  & 0.25  & 0.09  & 0.05  & 0.09  & 0.09  & 0.06 \\ [\cmsTabSkip]
Background (data) & 0.04  & 0.11  & 0.35  & 0.07  & 0.05  & 0.17  & 0.04 \\
Background (MC) & 0.01  & 0.29  & \NA  & 0.03  & 0.13  & \NA  & 0.07 \\ [\cmsTabSkip]
Method & 0.09  & 0.11  & 0.42  & 0.05  & 0.13  & 0.11  & 0.06 \\
Other & 0.07  & 0.12  & 0.24  & 0.02  & 0.10  & 0.03  & 0.06 \\ [\cmsTabSkip]
Total systematic & 1.31  & 1.04  & 1.21  & 0.74  & 0.82  & 1.02  & 0.41 \\
Statistical & 0.54  & 0.75  & 1.35  & 0.41  & 0.39  & 0.55  & 0.25 \\ [\cmsTabSkip]
Total & 1.42  & 1.28  & 1.82  & 0.84  & 0.91  & 1.15  & 0.48 \\
\end{scotch}
\end{table*}

\begin{table*}[!hp]
\centering
\topcaption{%
    Results and systematic uncertainties of the CMS \mt measurements, shown separately for the 7 and 8\TeV results in the dilepton (``dil''), lepton+jets (``lj''), and all-jets (``aj'') channels, for the 8\TeV results in the single top (``\PQt''), secondary vertex (``vtx''), and \PJGy analysis (``\PJGy''), and for their combination (``comb.'').
    All values are given in \GeVns.
}
\label{tab:data_input_CMS}
\renewcommand\arraystretch{1.2}
\cmsTable{\begin{scotch}{lrrr@{\hspace{4\tabcolsep}}rrrrrr@{\hspace{4\tabcolsep}}r}
& \multicolumn{10}{c}{CMS} \\
\cline{2-11}
& \multicolumn{3}{c@{\hspace{4\tabcolsep}}}{2011 (7\TeV)} & \multicolumn{6}{c@{\hspace{4\tabcolsep}}}{2012 (8\TeV)} & \multirow{2}{*}{comb.} \\
& \multicolumn{1}{c}{dil} & \multicolumn{1}{c}{lj} & \multicolumn{1}{c@{\hspace{4\tabcolsep}}}{aj} & \multicolumn{1}{c}{dil} & \multicolumn{1}{c}{lj} & \multicolumn{1}{c}{aj} & \multicolumn{1}{c}{\PQt} & \multicolumn{1}{c}{\PJGy} & \multicolumn{1}{c@{\hspace{4\tabcolsep}}}{vtx} & \\
\hline
\mt & 172.50  & 173.49  & 173.49  & 172.22  & 172.35  & 172.32  & 172.95  & 173.50  & 173.68  & 172.52 \\ [\cmsTabSkip]
\LHCJES1 & 0.77  & 0.24  & 0.69  & 0.31  & 0.10  & 0.16  & 0.40  & $<$0.01  & 0.11  & 0.06 \\
\LHCJES2 & 0.54  & 0.02  & 0.35  & 0.17  & 0.12  & 0.19  & 0.21  & $<$0.01  & 0.13  & 0.10 \\
\LHCJES3 & 0.06  & 0.01  & 0.08  & 0.03  & 0.01  & 0.02  & 0.05  & $<$0.01  & 0.01  & 0.01 \\
\LHCbJES & 0.70  & 0.61  & 0.49  & 0.37  & 0.32  & 0.29  & 0.38  & \NA  & \NA  & 0.25 \\
\LHCgJES & \NA  & \NA  & \NA  & 0.07  & 0.08  & 0.02  & \NA  & \NA  & \NA  & 0.04 \\
\LHClJES & \NA  & \NA  & \NA  & 0.04  & 0.06  & 0.01  & 0.07  & \NA  & \NA  & 0.05 \\
\CMSJES1 & 0.58  & 0.11  & 0.58  & \NA  & \NA  & \NA  & \NA  & \NA  & \NA  & 0.04 \\
JER & 0.14  & 0.23  & 0.15  & \NA  & 0.03  & 0.02  & 0.05  & $<$0.01  & 0.05  & 0.02 \\
Leptons & 0.14  & 0.02  & \NA  & 0.25  & 0.01  & \NA  & 0.05  & 0.10  & 0.24  & 0.07 \\
\PQb tagging & 0.09  & 0.12  & 0.06  & 0.01  & 0.06  & 0.02  & 0.10  & \NA  & 0.02  & 0.03 \\
\ptmiss & 0.12  & 0.06  & \NA  & 0.01  & 0.04  & \NA  & 0.15  & \NA  & \NA  & 0.01 \\
Pileup & 0.11  & 0.07  & 0.06  & 0.05  & 0.06  & 0.06  & 0.14  & 0.07  & 0.05  & 0.03 \\
Trigger & \NA  & \NA  & 0.24  & \NA  & \NA  & 0.01  & \NA  & 0.02  & \NA  & 0.01 \\ [\cmsTabSkip]
ME generator & 0.04  & 0.02  & 0.19  & 0.07  & 0.12  & 0.16  & \NA  & 0.37  & 0.42  & 0.14 \\
QCD radiation & 0.58  & 0.30  & 0.33  & 0.24  & 0.09  & 0.18  & 0.35  & 0.74  & 0.20  & 0.10 \\
Hadronization & \NA  & \NA  & \NA  & 0.38  & 0.01  & 0.04  & \NA  & 0.30  & 0.54  & 0.01 \\
CMS \PQb hadron \BR & \NA  & \NA  & \NA  & 0.12  & 0.16  & 0.13  & 0.15  & \NA  & 0.16  & 0.12 \\
Color reconnection & 0.13  & 0.54  & 0.15  & 0.13  & 0.01  & 0.16  & 0.05  & 0.12  & 0.08  & 0.03 \\
Underlying event & 0.05  & 0.15  & 0.20  & 0.11  & 0.08  & 0.14  & 0.20  & 0.13  & \NA  & 0.05 \\
PDF & 0.09  & 0.07  & 0.06  & 0.17  & 0.04  & 0.03  & 0.11  & 0.11  & 0.04  & $<$0.01 \\
CMS top quark \pt & \NA  & \NA  & \NA  & 0.51  & 0.02  & 0.06  & \NA  & \NA  & \NA  & 0.07 \\ [\cmsTabSkip]
Background (data) & \NA  & \NA  & 0.13  & \NA  & \NA  & 0.20  & \NA  & \NA  & 0.44  & 0.06 \\
Background (MC) & 0.05  & 0.13  & \NA  & \NA  & 0.03  & \NA  & 0.17  & 0.01  & \NA  & 0.01 \\ [\cmsTabSkip]
Method & 0.40  & 0.06  & 0.13  & \NA  & 0.04  & 0.06  & 0.39  & 0.22  & 0.62  & 0.09 \\
Other & \NA  & \NA  & \NA  & 0.03  & \NA  & \NA  & 0.25  & 0.09  & 0.09  & 0.01 \\ [\cmsTabSkip]
Total systematic & 1.52  & 0.97  & 1.23  & 0.94  & 0.45  & 0.57  & 0.93  & 0.94  & 1.11  & 0.39 \\
Statistical & 0.43  & 0.43  & 0.69  & 0.18  & 0.16  & 0.25  & 0.77  & 3.00  & 0.20  & 0.14 \\ [\cmsTabSkip]
Total & 1.58  & 1.06  & 1.41  & 0.95  & 0.48  & 0.62  & 1.20  & 3.14  & 1.12  & 0.42 \\
\end{scotch}}
\end{table*}

\begin{table*}[!hp]
\centering
\topcaption{%
    Pulls and weights of each input measurement in the LHC combination. The input measurements are the ATLAS and CMS 7 and 8\TeV \mt measurements in the dilepton (``dil''), lepton+jets (``lj''), and all-jets (``aj'') channels, and the CMS 8\TeV \mt measurements in the single top (``\PQt'), secondary vertex (``vtx''), and \PJGy analysis (``\PJGy'').
    The pull for measurement $i$ is defined as $(m_i - m_c)/\sqrt{\smash[b]{\sigma_i^2 - \sigma_c^2}}$,
    where $m_i$ ($\sigma_i$) is the central value (uncertainty) of the measurement and $m_c$ ($\sigma_c$) is the central value (uncertainty) of the LHC combination.
    The weights are rounded to two decimal places; when the full precision is used, the weights sum to one.
}
\label{tab:pulls_weights_LHC}
\renewcommand\arraystretch{1.2}
\cmsTable{\begin{scotch}{lccc@{\hspace{4\tabcolsep}}ccccccc@{\hspace{4\tabcolsep}}cccccc}
& \multicolumn{6}{c}{ATLAS} & & \multicolumn{9}{c}{CMS}  \\
\cline{2-7}\cline{9-17}
& \multicolumn{3}{c}{2011 (7\TeV)} & \multicolumn{3}{c}{2012 (8\TeV)} & & \multicolumn{3}{c}{2011 (7\TeV)} &  \multicolumn{6}{c}{2012 (8\TeV)} \\
& \multicolumn{1}{c}{dil} & \multicolumn{1}{c}{lj} & \multicolumn{1}{c@{\hspace{4\tabcolsep}}}{aj} & \multicolumn{1}{c}{dil} & \multicolumn{1}{c}{lj} & \multicolumn{1}{c}{aj} & & \multicolumn{1}{c}{dil} & \multicolumn{1}{c}{lj} & \multicolumn{1}{c@{\hspace{4\tabcolsep}}}{aj} & \multicolumn{1}{c}{dil} & \multicolumn{1}{c}{lj} & \multicolumn{1}{c}{aj} & \multicolumn{1}{c}{\PQt} & \multicolumn{1}{c}{\PJGy} & \multicolumn{1}{c}{vtx} \\
\hline
Pull & $+0.93$ & $-0.15$ & $+1.43$ & $+0.61$ & $-0.51$ & $+1.09$ & & $-0.01$ & $+0.96$ & $+0.71$ & $-0.33$ & $-0.47$ & $-0.37$ & $+0.38$ & $+0.31$ & $+1.08$ \\
Weight & $-0.02$ & $+0.07$ & $+0.00$ & $+0.16$ & $+0.17$ & $+0.03$ & & $-0.08$ & $-0.01$ & $+0.03$ & $+0.12$ & $+0.34$ & $+0.12$ & $-0.03$ & $+0.01$ & $+0.08$  \\
\end{scotch}}
\end{table*}

\begin{figure*}[!hp]
\centering
\includegraphics[width=\textwidth]{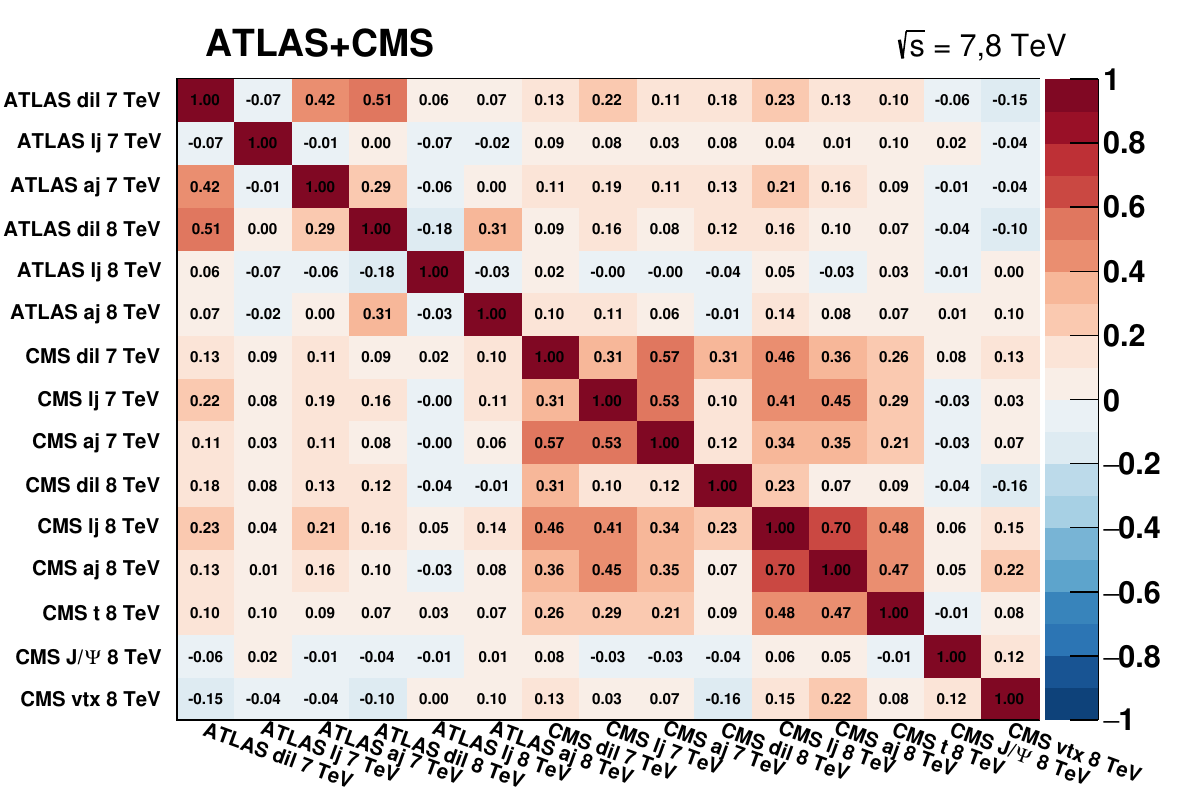}
\caption{%
    Correlation matrix for the ATLAS and CMS 7 and 8\TeV \mt measurements in the dilepton (``dil''), lepton+jets (``lj''), and all-jets (``aj'') channels, and for the CMS 8\TeV \mt measurements in the single top (``t''), secondary vertex (``vtx''), and \PJGy analysis (``\PJGy'').
}
\label{fig:LHC_corr}
\end{figure*}

\begin{table*}[!hp]
\centering
\topcaption{%
    Weights for each input measurement for the simultaneous combination of the four different channels. The input measurements are the ATLAS and CMS 7 and 8\TeV \mt measurements in the dilepton (``dil''), lepton+jets (``lj''), and all-jets (``aj'') channels, and the CMS 8\TeV \mt measurements in the single top (``t''), secondary vertex (``vtx''), and \PJGy analysis (``\PJGy''). The CMS alternative measurements are assigned to the ``Other'' channel.
}
\label{tab:weights_subcomb_ch}
\renewcommand\arraystretch{1.2}
\cmsTable{\begin{scotch}{lccc@{\hspace{4\tabcolsep}}ccccccc@{\hspace{4\tabcolsep}}cccccc}
& \multicolumn{6}{c}{ATLAS} & & \multicolumn{9}{c}{CMS}  \\
\cline{2-7}\cline{9-17}
& \multicolumn{3}{c}{2011 (7\TeV)} & \multicolumn{3}{c}{2012 (8\TeV)} & & \multicolumn{3}{c}{2011 (7\TeV)} &  \multicolumn{6}{c}{2012 (8\TeV)} \\
& \multicolumn{1}{c}{dil} & \multicolumn{1}{c}{lj} & \multicolumn{1}{c@{\hspace{4\tabcolsep}}}{aj} & \multicolumn{1}{c}{dil} & \multicolumn{1}{c}{lj} & \multicolumn{1}{c}{aj} & & \multicolumn{1}{c}{dil} & \multicolumn{1}{c}{lj} & \multicolumn{1}{c@{\hspace{4\tabcolsep}}}{aj} & \multicolumn{1}{c}{dil} & \multicolumn{1}{c}{lj} & \multicolumn{1}{c}{aj} & \multicolumn{1}{c}{\PQt} & \multicolumn{1}{c}{\PJGy} & \multicolumn{1}{c}{vtx} \\
\hline
ll & $+0.02$ & $+0.03$ & $-0.07$ & $+0.55$ & $+0.18$ & $-0.08$ & & $+0.10$ & $-0.02$ & $-0.07$ & $+0.33$ & $-0.19$ & $+0.22$ & $-0.08$ & $<$0.01 & $+0.08$ \\
lj & $-0.04$ & $+0.09$ & $+0.01$ & $+0.09$ & $+0.18$ & $+0.03$ & & $-0.10$ & $+0.03$ & $+0.03$ & $+0.05$ & $+0.71$ & $-0.06$ & $-0.06$ & $+0.01$ & $+0.06$ \\
aj & $-0.03$ & $+0.08$ & $+0.05$ & $+0.04$ & $+0.17$ & $+0.15$ & & $-0.13$ & $-0.13$ & $+0.13$ & $+0.12$ & $-0.12$ & $+0.67$ & $-0.05$ & $+0.01$ & $+0.04$ \\
Other & $+0.02$ & $+0.05$ & $+0.03$ & $+0.02$ & $+0.12$ & $+0.04$ & & $-0.18$ & $-0.04$ & $+0.10$ & $+0.14$ & $-0.12$ & $-0.18$ & $+0.46$ & $+0.05$ & $+0.49$ \\
\end{scotch}}
\end{table*}

\begin{table*}[!hp]
\centering
\topcaption{%
    Correlation matrix for the simultaneous combination of the dilepton (``dil''), lepton+jets (``lj''), all-jets (``aj''), and other (``Other'') channels.
}
\label{tab:corr_channels}
\renewcommand{\arraystretch}{1.2}
\begin{scotch}{crrrr}
& \multicolumn{1}{c}{dil} & \multicolumn{1}{c}{lj} & \multicolumn{1}{c}{aj} & \multicolumn{1}{c}{Other} \\ \hline
dil & 1.00 & 0.29 & 0.24 & $<$0.01 \\
lj & 0.29 & 1.00 & 0.59 & 0.31 \\
aj & 0.24 & 0.59 & 1.00 & 0.34 \\
Other & $<$0.01 & 0.31 & 0.34 & 1.00 \\
\end{scotch}
\end{table*}

}
\cleardoublepage \section{The CMS Collaboration \label{app:collab}}\begin{sloppypar}\hyphenpenalty=5000\widowpenalty=500\clubpenalty=5000
\cmsinstitute{Yerevan Physics Institute, Yerevan, Armenia}
{\tolerance=6000
A.~Hayrapetyan, A.~Tumasyan\cmsAuthorMark{1}\cmsorcid{0009-0000-0684-6742}
\par}
\cmsinstitute{Institut f\"{u}r Hochenergiephysik, Vienna, Austria}
{\tolerance=6000
W.~Adam\cmsorcid{0000-0001-9099-4341}, J.W.~Andrejkovic, T.~Bergauer\cmsorcid{0000-0002-5786-0293}, S.~Chatterjee\cmsorcid{0000-0003-2660-0349}, K.~Damanakis\cmsorcid{0000-0001-5389-2872}, M.~Dragicevic\cmsorcid{0000-0003-1967-6783}, P.S.~Hussain\cmsorcid{0000-0002-4825-5278}, M.~Jeitler\cmsAuthorMark{2}\cmsorcid{0000-0002-5141-9560}, N.~Krammer\cmsorcid{0000-0002-0548-0985}, A.~Li\cmsorcid{0000-0002-4547-116X}, D.~Liko\cmsorcid{0000-0002-3380-473X}, I.~Mikulec\cmsorcid{0000-0003-0385-2746}, J.~Schieck\cmsAuthorMark{2}\cmsorcid{0000-0002-1058-8093}, R.~Sch\"{o}fbeck\cmsorcid{0000-0002-2332-8784}, D.~Schwarz\cmsorcid{0000-0002-3821-7331}, M.~Sonawane\cmsorcid{0000-0003-0510-7010}, S.~Templ\cmsorcid{0000-0003-3137-5692}, W.~Waltenberger\cmsorcid{0000-0002-6215-7228}, C.-E.~Wulz\cmsAuthorMark{2}\cmsorcid{0000-0001-9226-5812}
% ...
\par}
\vskip\cmsinstskip
\dag:~Deceased\\
$^{1}$Also at Yerevan State University, Yerevan, Armenia\\
$^{2}$Also at TU Wien, Vienna, Austria\\
% ...

\newcommand{\AtlasOrcid}[2][]{\ifthenelse{\equal{#1}{}}{#2}{#2\cmsorcid{#1}}}
\section{The ATLAS Collaboration\label{app:atlas}}
\raggedright
\AtlasOrcid[0000-0002-6665-4934]{G.~Aad}$^\textrm{\scriptsize 102}$,
\AtlasOrcid[0000-0002-5888-2734]{B.~Abbott}$^\textrm{\scriptsize 120}$,
\AtlasOrcid[0000-0002-1002-1652]{K.~Abeling}$^\textrm{\scriptsize 55}$,
\AtlasOrcid[0000-0001-5763-2760]{N.J.~Abicht}$^\textrm{\scriptsize 49}$,
\AtlasOrcid[0000-0002-8496-9294]{S.H.~Abidi}$^\textrm{\scriptsize 29}$,
\AtlasOrcid[0000-0002-9987-2292]{A.~Aboulhorma}$^\textrm{\scriptsize 35e}$,
\AtlasOrcid[0000-0001-5329-6640]{H.~Abramowicz}$^\textrm{\scriptsize 151}$,
\AtlasOrcid[0000-0002-1599-2896]{H.~Abreu}$^\textrm{\scriptsize 150}$,
\AtlasOrcid[0000-0003-0403-3697]{Y.~Abulaiti}$^\textrm{\scriptsize 117}$,
\AtlasOrcid[0000-0002-8588-9157]{B.S.~Acharya}$^\textrm{\scriptsize 69a,69b,m}$
% ...
\bigskip
\\

% ...
$^{29}$Physics Department, Brookhaven National Laboratory, Upton NY; United States of America.\\
% ...
$^{35}$$^{(a)}$Facult\'e des Sciences Ain Chock, R\'eseau Universitaire de Physique des Hautes Energies - Universit\'e Hassan II, Casablanca;$^{(b)}$Facult\'{e} des Sciences, Universit\'{e} Ibn-Tofail, K\'{e}nitra;$^{(c)}$Facult\'e des Sciences Semlalia, Universit\'e Cadi Ayyad, LPHEA-Marrakech;$^{(d)}$LPMR, Facult\'e des Sciences, Universit\'e Mohamed Premier, Oujda;$^{(e)}$Facult\'e des sciences, Universit\'e Mohammed V, Rabat;$^{(f)}$Institute of Applied Physics, Mohammed VI Polytechnic University, Ben Guerir; Morocco.\\
% ...
$^{49}$Fakult\"{a}t Physik , Technische Universit{\"a}t Dortmund, Dortmund; Germany.\\
% ...
$^{55}$II. Physikalisches Institut, Georg-August-Universit\"{a}t G\"ottingen, G\"ottingen; Germany.\\
% ...
$^{69}$$^{(a)}$INFN Gruppo Collegato di Udine, Sezione di Trieste, Udine;$^{(b)}$ICTP, Trieste;$^{(c)}$Dipartimento Politecnico di Ingegneria e Architettura, Universit\`a di Udine, Udine; Italy.\\
% ...
$^{102}$CPPM, Aix-Marseille Universit\'e, CNRS/IN2P3, Marseille; France.\\
% ...
$^{117}$Department of Physics, New York University, New York NY; United States of America.\\
% ...
$^{120}$Homer L. Dodge Department of Physics and Astronomy, University of Oklahoma, Norman OK; United States of America.\\
% ...
$^{150}$Department of Physics, Technion, Israel Institute of Technology, Haifa; Israel.\\
$^{151}$Raymond and Beverly Sackler School of Physics and Astronomy, Tel Aviv University, Tel Aviv; Israel.\\
% ...
$^{a}$ Also Affiliated with an institute covered by a cooperation agreement with CERN.\\
$^{b}$ Also at An-Najah National University, Nablus; Palestine.\\
$^{c}$ Also at Borough of Manhattan Community College, City University of New York, New York NY; United States of America.\\
$^{d}$ Also at Center for High Energy Physics, Peking University; China.\\
$^{e}$ Also at Center for Interdisciplinary Research and Innovation (CIRI-AUTH), Thessaloniki; Greece.\\
$^{f}$ Also at Centro Studi e Ricerche Enrico Fermi; Italy.\\
% ...
$^{m}$ Also at Department of Physics, King's College London, London; United Kingdom.\\
% ...
\end{sloppypar}
%%% END EDITABLE REGION %%%
% skeleton_end
\end{document}